\documentclass[reprint,amsmath,amssymb,aps,showkeys]{revtex4-2}
\usepackage{siunitx}
\usepackage{color}
\usepackage{graphicx}
\usepackage{dcolumn}
\usepackage{bm}
\usepackage[dvipsnames]{xcolor}

\begin{document}

\preprint{APS/123-QED}

\title{Rare event-triggered transitions in aerodynamic bifurcation}

\author{Ariane Gayout}
\author{Micka\"el Bourgoin}
\author{Nicolas Plihon}
\affiliation{Univ Lyon, ENS de Lyon, Univ Claude Bernard Lyon 1, CNRS, Laboratoire de Physique, F-69342 Lyon, France}

\date{\today}
             
\begin{abstract}
The transitions between two states of a bistable system are investigated experimentally and analyzed in the framework of rare-event statistics. Considering a disk pendulum swept by a flow in a wind tunnel, bistability between two aerodynamic branches is observed, with spontaneous transitions from one branch to the other. The waiting times before spontaneous transition are distributed following a double-exponential as a function of the control parameter, spanning four orders of magnitude in time, for both transitions. Inspired by a model originally applied to the transition to turbulence, we show that, for the disk pendulum, both transitions are controlled by rare events of the aerodynamic forces acting on the disk which we propose to link in particular to the vortex shedding-induced fluctuations. Beyond the aerodynamic aspects, this work has interesting fundamental outcomes regarding the broad field of rare events in out-of-equilibrium systems.

\end{abstract}

\pacs{Valid PACS appear here}

\maketitle

\emph{Introduction}.---
Turbulent flows are canonical out-of-equilibrium random systems in which transitions between large-scale modes can be triggered by rare events. Such phenomena have been investigated in laboratory experiments, including thermal convection~\cite{bib:sreeni2002_PRE,bib:sugiyama2010_PRL,bib:mishra2011_JFM}, Couette flows~\cite{Zimmerman2011}, von K\'arm\'an swirling flows~\cite{Douady1991,Ravelet2004,bib:deLaTorre2007_PRL}, rotating turbulence~\cite{tian2001}, or the dynamo instability~\cite{Faranda2014}. It is also observed in the atmosphere~\cite{Schmeits2001,Baldwin2001,bib:weeks1997_Science}, with dramatic consequences on climate, or in industrial systems~\cite{Ahmed1984,Goman1985,Cadot2015,Evrard2016, Bonnavion2018,bib:haffner2020_JFM}, with implications on energy efficiency and fatigue of the structures. In spite of their variety, these systems share the feature of exhibiting spontaneous transitions between multi-stable large-scale modes, with waiting times ahead of transitions exponentially distributed~\cite{Hof2006,bib:deLaTorre2007_PRL,bib:sreeni2002_PRE,Ravelet2004,bib:sugiyama2010_PRL,bib:mishra2011_JFM}, and characteristic times several orders of magnitude larger than any hydrodynamic time of the underlying flow. This places such transitions in the field of rare events. Theoretical approaches inspired from statistical thermodynamics accurately predict the multi-stable large-scale modes, imposed by conservation laws and symmetry considerations~\cite{bib:robert1991_JFM,bib:miller1992_PRA,bib:naso2010_PRE, Bouchet2012}. Large deviation theory~\cite{bib:vulpiani2014_book} offers a framework to compute rare events of climate dynamics~\cite{Ragone2018} or the turbulent wake of a body immersed in a flow~\cite{Lestang2018}. An alternative approach is related to the computation of instantons~\cite{Falkovich1996}, with verifications in experiments~\cite{Dematteis2019} and direct numerical simulations~\cite{Grafke2015}.

Among the variety of bifurcations observed in turbulent flows, the long standing question of the transition to turbulence in pipe flows, first introduced by Reynolds more than a century ago \cite{Reynolds1883}, was only recently elucidated~\cite{Barkley2016}. Recent thorough experiments showed that the transition is triggered by local fluctuations of the turbulence intensity, within localized intermittent puffs~\cite{Hof2006}. The lifetime of turbulent puffs and their splitting time are distributed following a double exponential as a function of the Reynolds number, and the critical Reynolds number is defined from equiprobability~\cite{Hof2006,Avila2011}.
This led to the development of a simple model linking the observed statistics for the transition to turbulence to the phenomenology of extreme fluctuations of the kinetic energy within the localized turbulent puffs~\cite{Goldenfeld2010}.
Numerous careful numerical investigations legitimated this framework of rare-event dynamics~\cite{Nemoto2018,Schikarski2019, Nemoto2020}. In this Letter, we present a simple canonical configuration to experimentally investigate rare events triggered transition in out-of-equilibrium systems: a simple pendular disk subject to a flow displays a well-characterized bi-stable dynamics~\cite{Obligado2013}, for which transitions occur as the pendulum interacts with the flow.  These spontaneous transitions exhibit rare-event statistics, successfully described in the framework introduced in Ref~\cite{Goldenfeld2010} for pipe flows. We propose a phenomenology based on the extreme fluctuations of the aerodynamic torque. From the fundamental point of view, this study brings a simple configuration for a systematic investigation of rare-event statistics, and adds up to previous observations of double-exponential distributions of turbulent lifetimes~\cite{Linkmann2015,Shi2013,Gome2020}.

\begin{figure}[!ht]
\includegraphics[width=\linewidth]{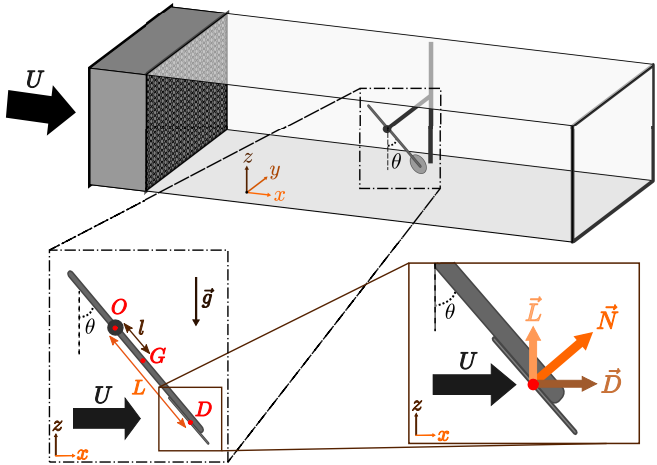}
\caption{Experimental setup showing the test section of the wind-tunnel and the pendulum made of a thin disk. See text for details. 
\label{fig:scheme}}
\end{figure}

\emph{Experimental setup}.---
A sketch of the the experimental setup is provided in Fig.\ref{fig:scheme}. A pendulum made of a thin aluminium disk of surface $ S = 13\, {\rm cm}^2$  is fixed at the end of a sanded saw blade of length \SI{30}{\centi\meter}. The pendulum is swept by a flow in a wind tunnel, and is free to rotate around the horizontal axis in the span-wise direction, at point $O$. The angular position $\theta$ with respect to the vertical is recorded from a friction-less potentiometer. The center of mass $G$ of the pendulum is such that $OG=l=\SI{5.7}{\centi\meter}$, and the center of the disk $D$, where the aerodynamic forces acting on the disk apply, is such that $OD=L=\SI{19.5}{\centi\meter}$. The total mass of the pendulum is $m=\SI{17.01}{\gram}$. Here, we neglect the aerodynamic forces acting on the blade and the influence of the disk thickness (\SI{0.3}{\milli\meter}) on the aerodynamic forces exerted by the flow on the disk. 
The wind tunnel is a closed channel with a square cross-section of size \SI{51}{\centi\meter}. Before impacting the pendulum, the flow is conditioned through a \SI{6}{\milli\meter}-diameter honeycomb. 
The turbulence rate is constant for all flow velocities $U$ and is equal to $1.5\%$. The physical control parameter of the experiment is the dynamic pressure $\rho U^2$ (which controls the amplitude of the aerodynamic forces acting on the disk), with $\rho$ the density of air, while our experimental control parameter is the rotation rate of the wind turbine generating the flow, and thus the velocity $U$.  

The aerodynamic forces exerted by the flow on the disk create an aerodynamic torque $\Gamma_{aero}(t)$ at point $O$ and the equation of motion reads:
\begin{equation}\label{eq:dyn}
J \ddot{\theta}=-m g l \mathrm{sin}(\theta) + \Gamma_{aero}(t), 
\end{equation}
where $J$ is the moment of inertia of the pendulum.

In the steady state regime, the flow creates a drag force $\mathbf{D}= \frac{1}{2}C_D \rho U^2 S$ and a lift force $\mathbf{L}= \frac{1}{2}C_L \rho U^2 S$ on the disk (see Fig.~\ref{fig:scheme}), with a drag coefficient $C_D$ and a lift coefficient $C_L$. These aerodynamic forces result in a time-averaged aerodynamic torque at point $O$ expressed as $\frac{1}{2}\rho L S U^2 C_N(\theta)$ where the \textit{normal coefficient} $C_N$ is the  projection of $C_D$ and $C_L$, along the direction normal to the disk. Note that the aerodynamic coefficients $C_D$, $C_L$ and $C_N$ depend on the angle $\theta$~\cite{Flachsbart1932}. In the steady state regime, the torque induced by the weight of the pendulum balances the aerodynamic torque:
\begin{equation}\label{eq:start}
- \Gamma_{weight}=m g l \sin(\theta) = \frac{1}{2}\rho L S U^2 C_N(\theta) = \Gamma_{aero}.
\end{equation}


\begin{figure}[!ht]
\includegraphics[width=\linewidth]{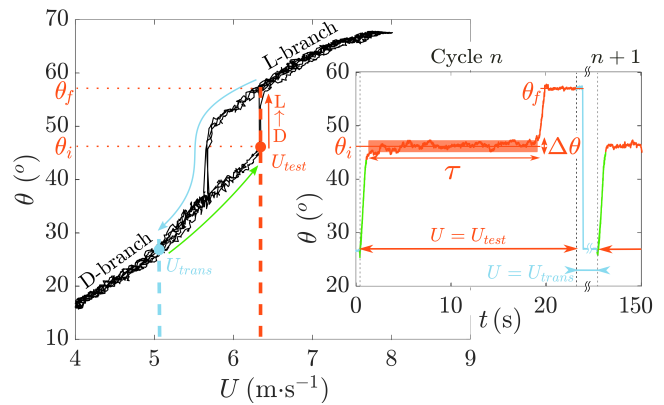}
\caption{Left: Angular position $\theta$ of the pendulum as a function of $U$. Right: time series illustrating the protocol used to probe the statistics of spontaneous jumps from the the D- to the L-branch, at $U_{\rm test}=$ \SI{6.4}{\meter\per\second} (see text for details).\label{fig:definition}}
\end{figure}

\emph{Hysteresis and spontaneous transitions.}--- The black curve on the left panel of Fig.~\ref{fig:definition} shows the angular position of the pendulum as a function of the wind velocity, clearly evidencing the subcritical bifurcation between two equilibrium branches~\cite{Obligado2013}.
This curve was obtained by slowly sweeping the wind velocity over the full range of interest at a low frequency (\SI{1}{\milli\hertz}),  preventing dynamic stall~\cite{Ericsson1980, McCroskey1975}.  
 The lower (resp. upper) branch is referred to as the D (resp. L)-branch, as related to the predominance of drag or lift in the torque balance. Increasing $U$ from low velocities, the pendulum stays in the D-branch, suddenly jumps in the L-branch at $U=\SI{6.4}{\meter\per\second}$ (for the present experimental parameters) and stays on the L-branch as long as the velocity is not decreased below \SI{5.7}{\meter\per\second}. The origin of the bistability was shown to be linked to the sharp stall of $C_N(\theta)$ at $\theta\simeq\SI{50}{\degree}$~\cite{Obligado2013}. In the remaining of this Letter, the transitions in the upward (resp. downward) direction will be referred to as D$\rightarrow$L (resp. L$\rightarrow$D) transitions. 
In order to probe the statistical properties of the transitions, we developed a  specific  experimental  protocol, sketched in the right panel of Fig.~\ref{fig:definition} for the case of D$\rightarrow$L transitions. Indeed, as the system is bistable, when the pendulum jumps from the D to the L-branch at $U=\SI{6.4}{\meter\per\second}$, it stays in the L-branch as long as the control parameter is kept constant; in our protocol the flow velocity is modulated in order to aggregate the statistics of thousands of transitions. At time $t=0$, the wind velocity, initially at $U_{trans}=\SI{5}{\meter\per\second}$, below the bistable region, is set to a fixed test velocity $U_{test}$,within the bistable region (e.g. \SI{6.4}{\meter\per\second}) . After a short transient (dynamics displayed in green in Fig.~\ref{fig:definition}), the pendulum reaches the initial average equilibrium angle $\theta_i$ (in the D-branch), and spontaneously jumps to the L-branch after a time $\tau$, finally reaching a final average equilibrium angle $\theta_f$ (displayed in orange in Fig.~\ref{fig:definition}). A few seconds after the transition, the flow velocity is decreased back to $U_{trans}$ where it is maintained for typically two minutes in order to restore the global flow structure in the wind tunnel (displayed in light blue in Fig.~\ref{fig:definition}) before repeating the cycle. During the time interval $\tau$, the equilibrium position of the pendulum fluctuates around the average equilibrium angle $\theta_i$, with a standard deviation $\Delta\theta$. Note that $\Delta \theta \ll |\theta_f-\theta_i|$, so that there is no ambiguity between the occurrence of a transition and simple natural fluctuations. This protocol is repeated hundreds of times (typically 200 times) for several values of the test velocities $U_{test}$, allowing to analyse the statistics of the waiting times $\tau$ as a function of the control parameter of the bifurcation. A similar protocol is used for probing the L$\rightarrow$D transition, for which $U_{trans}$ lies in the L-branch, above the bistable region (e.g. \SI{7}{\meter\per\second}), and is slowly decreased down to a prescribed test velocity  $U_{test}$ within the bistable region (e.g. \SI{5.7}{\meter\per\second}).

\begin{figure}[!ht]
\includegraphics[width=\linewidth]{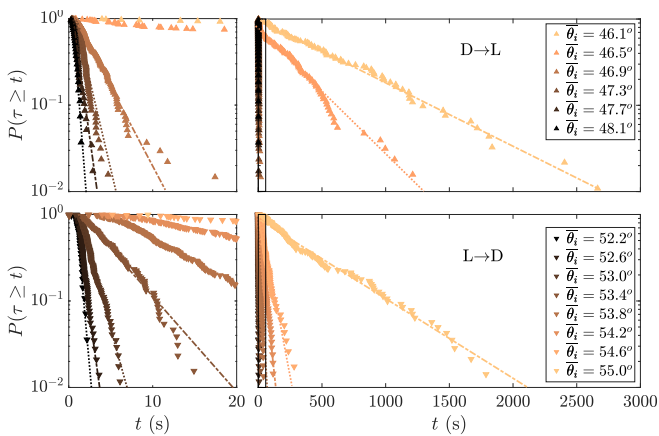}
\caption{Cumulative distribution function of the waiting time $\tau$ for different values of $\overline{\theta_i}$ for the (top)  D$\rightarrow$L and (bottom) L$\rightarrow$D  transition. Dotted lines represent the exponential fits.\label{fig:stat}}
\end{figure}

The above-detailed protocol gives access to the statistical distribution of the waiting times $\tau$ as a function the test velocity $U_{test}$, simply referred to as $U$ in the sequel for clarity. Note that Eq.~\ref{eq:start} shows that the actual physical control parameter controlling the evolution of $\theta$ is the dynamic pressure $\rho U^2$, as previously mentioned. We thus chose to present our experimental results as a function of the initial angle $\theta_i$, which is a proxy for the physical control parameter. Because of small variations of atmospheric pressure, and hence air density $\rho$, a series of measurements at constant velocity $U$ results in a series of $\theta_i$ values typically spanning \SI{0.14}{\degree} around a mean value $\overline{\theta_i}$, from which the series are labeled. 
For both the D$\rightarrow$L and the  L$\rightarrow$D transitions, the cumulative density function of waiting times $\tau$ (Fig.~\ref{fig:stat}) shows that the probability for the pendulum to undergo a transition after a given time $t$ follows an exponential law: $P_{\overline{\theta_i}}(\tau\geq t)=\exp(-t/\tau_c(\overline{\theta_i}))$, as observed in other multi-stable systems~\cite{Hof2006,bib:deLaTorre2007_PRL,bib:sreeni2002_PRE,Ravelet2004,bib:sugiyama2010_PRL,bib:mishra2011_JFM,Avila2010}. A striking feature is that the characteristic time-scale $\tau_c$ strongly depends on $\overline{\theta_i}$: it spans nearly four orders of magnitude when $\overline{\theta_i}$ spans only a few degrees. 

A closer look at Fig.~\ref{fig:stat} shows deviations from the exponential law for some experiments, in particular in the long term behavior (e.g. $ \overline{\theta_i} =\SI{46.9}{\degree}$ for the D$\rightarrow$L  transition)
 --  likely due to finite size sampling, as observed for similar distributions \cite{Avila2010} or small deviations of the physical control parameter $\rho U^2$ during a series of measurements at constant velocity $U$ due to meteorological variations.

\begin{figure}[!ht]
 \includegraphics[width=\linewidth]{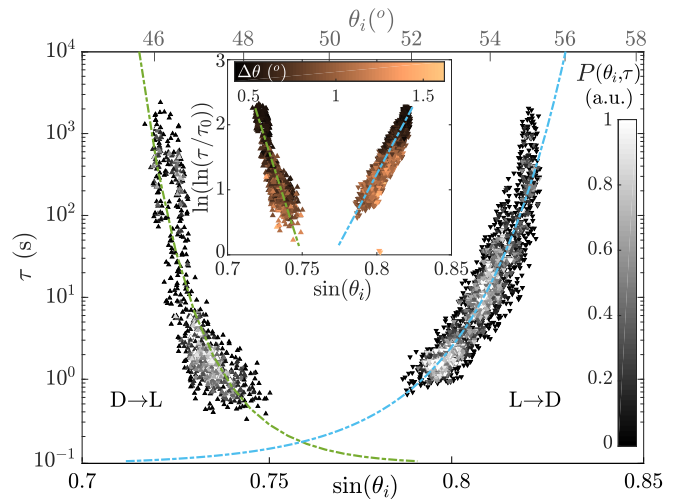}
 \caption{Waiting time $\tau$ as a function of $\sin(\theta_i)$ for both transitions. Dash-dotted corresponds to the best fits according to Eq.~\ref{eq:supexp}. Inset: Evolution of  $\ln(\ln(\tau/\tau_0))$ as a function of $\sin(\theta_i)$. See text for details.\label{fig:eagle1}}
 \end{figure}

\emph{Double-exponential distribution}.---
We will show in the remaining of this Letter that the dynamics of the spontaneous transitions is governed by rare fluctuations of the aerodynamic torque $\Gamma_{aero}$, which, according to Eq. \ref{eq:start}, is proportional to $\sin(\theta)$. 
Figure~\ref{fig:eagle1} thus shows all individual events $\tau$ for the 2596 recorded spontaneous transitions as a function of $\sin(\theta_i)$.
The colorbar of Fig.~\ref{fig:eagle1} corresponds to the joint  probability of the variables $\sin(\theta_i)$ and $\tau$, computed as the local density of the experimental points in the $(\sin(\theta_i),\tau)$ space. As expected, the characteristic times $\tau_c(\overline{\theta_i})$ computed from the exponential fits for several values of $\overline{\theta_i}$ are located close to the brightest spots in Fig.~\ref{fig:eagle1} (not displayed for legibility).

Weighting the experimental points with their joint probability, the distribution of waiting times for each transition $T$ (D$\rightarrow$L or L$\rightarrow$D) is fitted with excellent agreement following a double-exponential evolution as:
\begin{equation}\label{eq:supexp}
    \tau=\tau_0^T\exp\left[\exp\left(\frac{\sin(\theta_i)-\sin(\theta_0^T)}{\eta^T}\right)\right].
\end{equation}

Note that in the fitting procedure, only $\sin(\theta_0^T)$ and $\eta^T$ are free fitting parameters, while $\tau_0^T$ (which, according to previous studies reporting similar double exponential statistics, is expected to be a characteristic time scale of the problem) is taken based on the spectral signature of the pendulum fluctuations as the frequency inverse of the first peak in the power spectral density of $\theta$ (see Supplementary Fig. 1). The best fits are shown as dash-dotted lines in Fig.~\ref{fig:eagle1} for both transitions.

\begin{figure}[!ht]
 \includegraphics[width=\linewidth]{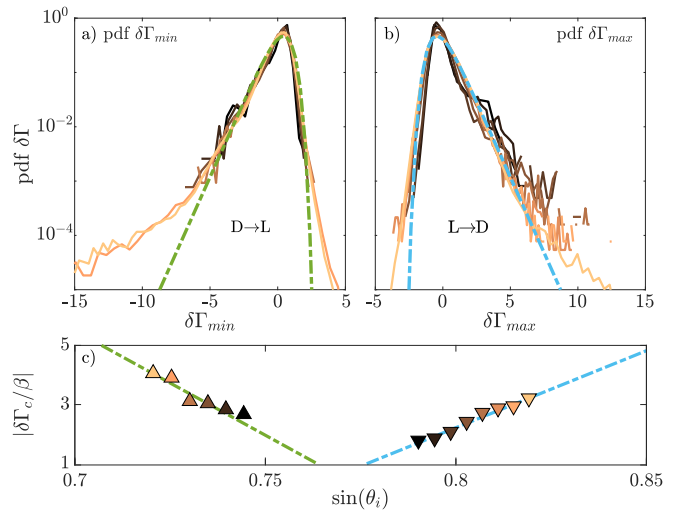}%
 \caption{Normalized and centered pdf of the torque-fluctuation a)  minima $\delta\Gamma_{min}$ and b) maxima $\delta\Gamma_{max}$ and associated fitted Gumbel distributions (dash-dotted lines). c) Evolution of $\delta\Gamma_c/\beta$, as a function of $\sin(\overline{\theta_i})$ and linear fit (dash-dotted lines). Color code for $\overline{\theta_i}$ identical to Fig.~\ref{fig:stat}.\label{fig:seuil}}
 \end{figure}

Similar double-exponential statistics were reported for the characteristic lifetime of turbulent puffs as a function of the Reynolds number Re during the transition to turbulence in pipe flows~\cite{Hof2008,Barkley2016}
and were analyzed in the framework of rare-event dynamics~\cite{Goldenfeld2010,Nemoto2018,Nemoto2020}. 
In the sequel, we analyze the statistics of the waiting times of the pendulum in a similar framework. 
Briefly, for pipe flows, the phenomenology assumes that the turbulent state cannot be sustained when the turbulent kinetic energy lies below a given threshold ; hence turbulence dies if all local maxima of the kinetic energy lie below this threshold. Maxima of the turbulent kinetic energy follow a Gumbel distribution, leading to a double-exponential statistics for the lifetime of turbulent puffs. Let us extend this approach to the bistable pendulum by focusing first on the L$\rightarrow$D transition, for which we propose the following mechanism: the pendulum jumps when the L state cannot be sustained, or equivalently when all maxima of the aerodynamic torque lie below a given threshold.  The aerodynamic torque $\Gamma_{aero}(t)$ is computed from the time series $\theta(t)$ following Eq.~\ref{eq:dyn}.  The probability density functions (pdf) of the torque fluctuations $\delta\Gamma = \Gamma_{aero}(t)-\langle\Gamma_{aero}(t)\rangle$ (where $\langle\Gamma_{aero}(t)\rangle$ is the time-averaged torque) display exponential tails (see Supplementary Fig. 2), allowing the application of rare-event statistics. The pdf of the maxima $\delta\Gamma_{max}$ of $\delta\Gamma$, computed over time intervals $\tau_0^{\rm L \rightarrow D}$ for each $\overline{\theta_i}$, are displayed in Fig~\ref{fig:seuil} b)). They are in excellent agreement with Gumbel distributions 
$ P(\delta\Gamma_{max})=\frac{1}{\beta}\exp\left[-X/\beta +\exp\left(-X/\beta\right)\right]$,
where $X = \delta\Gamma_{max} -\mu$ and $\mu$ and $\beta$ are calculated from the mean and the standard deviation of the distribution~\footnote{The mean value being $\mu+\beta\gamma$ , with $\gamma\simeq 0.577$ the Euler-Mascheroni constant and the standard deviation being $\beta\pi/\sqrt{6}$. For each value of $\overline{\theta_i}$, $\beta$ and $\mu$ have been extracted from experimental signals.}, as shown in Fig.~\ref{fig:seuil} (and further checked on the cumulative distribution function in Supplementary Fig.3).
Assuming that a L$\rightarrow$D transition occurs when the maximum $\delta\Gamma_{max}$ of torque fluctuations lies below a threshold $\delta\Gamma_c$ during a time $\tau_0^{\rm L \rightarrow D}$, we now apply the statistical model developed for turbulent puffs~\cite{Goldenfeld2010} to our experimental data in order to compute $\delta\Gamma_c$ as a function of $\sin(\overline{\theta_i})$. For a given value of $\overline{\theta_i}$, the probability $p$ that the transition occurs during $\tau_0^{{\rm L}\rightarrow {\rm D}}$ is given as $p= \tau_0^{{\rm L}\rightarrow {\rm D}}/\tau_c(\overline{\theta_i})$, and is linked to the pdf of the maxima of $\delta\Gamma$ as $p = P(\delta\Gamma_{max}<\delta\Gamma_c)$.

Using the parameters $\beta$ and $\mu$ extracted from the pdf of  $\delta\Gamma_{max}$, the threshold $\delta\Gamma_c$ can then be estimated from the relation~\cite{Goldenfeld2010}:
\begin{equation}\label{eq:seuil}
p(\overline{\theta_i}) = \frac{\tau_0}{\tau_c(\overline{\theta_i})} = \exp\left[-\exp\left(-\frac{\delta\Gamma_c(\overline{\theta_i})-\mu(\overline{\theta_i})}{\beta(\overline{\theta_i})}\right)\right]. 
\end{equation}

The evolution of $\delta\Gamma_c/\beta$ as a function of $\sin(\overline{\theta_i})$ is shown in Fig.~\ref{fig:seuil} c). The best linear regression (dash-dotted lines) matches the slope of the double-exponential fit following Eq.~\ref{eq:supexp}, and, more generally, the linear evolution is understood from the fact that $\mu(\overline{\theta_i})/\beta(\overline{\theta_i})$ is observed constant for the L$\rightarrow$D transition (see Supplementary Fig. 4). As $\sin(\overline{\theta_i})$ decreases from 0.82 to 0.79, the threshold normalized to the torque standard deviation decreases, which corresponds to more probable transitions, and thus shorter waiting times.
 On the other hand, the standard deviation of the torque fluctuations strongly increases as $\sin(\overline{\theta_i})$ decreases  (see Supplementary Fig. 4), which is reminiscent of the increase of standard deviation of the pendulum angular position $\Delta\theta$, as color-coded in the inset of Fig.~\ref{fig:eagle1}).

The analysis proposed for the L$\rightarrow$D transition also applies to the D$\rightarrow$L transition. A transition occurs when the D state cannot be sustained, or equivalently when the minimum aerodynamic torque during time $\tau_0^{\rm D\rightarrow L}$ lies above a given threshold. This transition is thus controlled by the pdf of the minima $\delta\Gamma_{min}$ of $\delta\Gamma$, computed over $\tau_0^{\rm D\rightarrow L}$, and displayed in Fig.~\ref{fig:seuil} a) to be in excellent agreement with Gumbel distributions. A similar application of the rare-event statistical model leads to conclusions similar to the ones drawn for the L$\rightarrow$D transition and summarized in Fig.~\ref{fig:seuil} c) for $\sin(\overline{\theta_i})$ between 0.725 and 0.745 (see also Supplementary Fig. 4). Note that, for both transitions, the linear best fits of $\delta\Gamma_c/\beta$ as a function of $\sin(\overline{\theta_i})$ shown as dash-dotted lines in Fig.~\ref{fig:seuil} c) are in excellent agreement with those shown in the inset of Fig.~\ref{fig:eagle1}, a strong asset for the validity of the proposed phenomenology.

\emph{Conclusion.---} We have presented a simple out-of-equilibrium bistable system which exhibits canonical rare-event statistics which were successfully interpreted under the light of a phenomenological model inspired by previous studies on the transition to turbulence. This work extends the previous approach to the case of turbulent wakes, what may pave the way towards the modeling of other situations (transition between large-scale modes in Rayleigh-B\'enard, zonal flows, swirling flows, etc.). However, the origin of the aerodynamic fluctuations triggering the transitions requires further investigations, that require time-resolved 3D visualization of the structure of the flow around the pendulum, a task well-beyond the scope of the present study. Indeed, the aerodynamic torque applied to the pendulum strongly depends upon the structure of the boundary layer upstream and downstream of the disk, controlled by vortex shedding. The characteristic times $\tau_0^T$ identified in the present work correspond to Strouhal numbers $St \simeq 0.07$ and recall values reported in previous studies for transverse vortex shedding (in the $x-y$ plane) for a disk at non zero incidence~\cite{Gao2018}. We thus conjecture that transverse vortex shedding controls the structure of the boundary layer and thus of rare events of the aerodynamic torque. This would point toward a leading role of transverse vortices, with potential important implications on a broader range of fluid-structure applications, as for instance such rare transitions may therefore strongly depend on the aspect ratio of the immersed body.
Finally, the simplicity of implementation of the experimental system together with the generic observations reported on its rare-event behavior, makes it an ideal configuration to test theoretical models for out-of-equilibrium bistable systems, crucial for major applications, such as climate change.

\acknowledgements{This work was supported by Initiative d’Excellence de Lyon (IDEXLYON) of the University of Lyon in the framework of the Programme Investissements d’Avenir (ANR-16- IDEX-0005).}
\bibliographystyle{apsrev4-1}
\bibliography{ref_article1}

\begin{thebibliography}{47}%
\makeatletter
\providecommand \@ifxundefined [1]{%
 \@ifx{#1\undefined}
}%
\providecommand \@ifnum [1]{%
 \ifnum #1\expandafter \@firstoftwo
 \else \expandafter \@secondoftwo
 \fi
}%
\providecommand \@ifx [1]{%
 \ifx #1\expandafter \@firstoftwo
 \else \expandafter \@secondoftwo
 \fi
}%
\providecommand \natexlab [1]{#1}%
\providecommand \enquote  [1]{``#1''}%
\providecommand \bibnamefont  [1]{#1}%
\providecommand \bibfnamefont [1]{#1}%
\providecommand \citenamefont [1]{#1}%
\providecommand \href@noop [0]{\@secondoftwo}%
\providecommand \href [0]{\begingroup \@sanitize@url \@href}%
\providecommand \@href[1]{\@@startlink{#1}\@@href}%
\providecommand \@@href[1]{\endgroup#1\@@endlink}%
\providecommand \@sanitize@url [0]{\catcode `\\12\catcode `\$12\catcode
  `\&12\catcode `\#12\catcode `\^12\catcode `\_12\catcode `\%12\relax}%
\providecommand \@@startlink[1]{}%
\providecommand \@@endlink[0]{}%
\providecommand \url  [0]{\begingroup\@sanitize@url \@url }%
\providecommand \@url [1]{\endgroup\@href {#1}{\urlprefix }}%
\providecommand \urlprefix  [0]{URL }%
\providecommand \Eprint [0]{\href }%
\providecommand \doibase [0]{http://dx.doi.org/}%
\providecommand \selectlanguage [0]{\@gobble}%
\providecommand \bibinfo  [0]{\@secondoftwo}%
\providecommand \bibfield  [0]{\@secondoftwo}%
\providecommand \translation [1]{[#1]}%
\providecommand \BibitemOpen [0]{}%
\providecommand \bibitemStop [0]{}%
\providecommand \bibitemNoStop [0]{.\EOS\space}%
\providecommand \EOS [0]{\spacefactor3000\relax}%
\providecommand \BibitemShut  [1]{\csname bibitem#1\endcsname}%
\let\auto@bib@innerbib\@empty
\bibitem [{\citenamefont {Sreenivasan}\ \emph {et~al.}(2002)\citenamefont
  {Sreenivasan}, \citenamefont {Bershadskii},\ and\ \citenamefont
  {Niemela}}]{bib:sreeni2002_PRE}%
  \BibitemOpen
  \bibfield  {author} {\bibinfo {author} {\bibfnamefont {K.~R.}\ \bibnamefont
  {Sreenivasan}}, \bibinfo {author} {\bibfnamefont {A.}~\bibnamefont
  {Bershadskii}}, \ and\ \bibinfo {author} {\bibfnamefont {J.~J.}\ \bibnamefont
  {Niemela}},\ }\href {\doibase 10.1103/PhysRevE.65.056306} {\bibfield
  {journal} {\bibinfo  {journal} {Phys. Rev. E}\ }\textbf {\bibinfo {volume}
  {65}},\ \bibinfo {pages} {056306} (\bibinfo {year} {2002})}\BibitemShut
  {NoStop}%
\bibitem [{\citenamefont {Sugiyama}\ \emph {et~al.}(2010)\citenamefont
  {Sugiyama}, \citenamefont {Ni}, \citenamefont {Stevens}, \citenamefont
  {Chan}, \citenamefont {Zhou}, \citenamefont {Xi}, \citenamefont {Sun},
  \citenamefont {Grossmann}, \citenamefont {Xia},\ and\ \citenamefont
  {Lohse}}]{bib:sugiyama2010_PRL}%
  \BibitemOpen
  \bibfield  {author} {\bibinfo {author} {\bibfnamefont {K.}~\bibnamefont
  {Sugiyama}}, \bibinfo {author} {\bibfnamefont {R.}~\bibnamefont {Ni}},
  \bibinfo {author} {\bibfnamefont {R.~J.}\ \bibnamefont {Stevens}}, \bibinfo
  {author} {\bibfnamefont {T.~S.}\ \bibnamefont {Chan}}, \bibinfo {author}
  {\bibfnamefont {S.~Q.}\ \bibnamefont {Zhou}}, \bibinfo {author}
  {\bibfnamefont {H.~D.}\ \bibnamefont {Xi}}, \bibinfo {author} {\bibfnamefont
  {C.}~\bibnamefont {Sun}}, \bibinfo {author} {\bibfnamefont {S.}~\bibnamefont
  {Grossmann}}, \bibinfo {author} {\bibfnamefont {K.~Q.}\ \bibnamefont {Xia}},
  \ and\ \bibinfo {author} {\bibfnamefont {D.}~\bibnamefont {Lohse}},\ }\href
  {\doibase 10.1103/PhysRevLett.105.034503} {\bibfield  {journal} {\bibinfo
  {journal} {Phys. Rev. Lett.}\ }\textbf {\bibinfo {volume} {105}},\ \bibinfo
  {pages} {034503} (\bibinfo {year} {2010})}\BibitemShut {NoStop}%
\bibitem [{\citenamefont {Mishra}\ \emph {et~al.}(2011)\citenamefont {Mishra},
  \citenamefont {De}, \citenamefont {Verma},\ and\ \citenamefont
  {Eswaran}}]{bib:mishra2011_JFM}%
  \BibitemOpen
  \bibfield  {author} {\bibinfo {author} {\bibfnamefont {P.~K.}\ \bibnamefont
  {Mishra}}, \bibinfo {author} {\bibfnamefont {A.~K.}\ \bibnamefont {De}},
  \bibinfo {author} {\bibfnamefont {M.~K.}\ \bibnamefont {Verma}}, \ and\
  \bibinfo {author} {\bibfnamefont {V.}~\bibnamefont {Eswaran}},\ }\href
  {\doibase 10.1017/S0022112010004830} {\bibfield  {journal} {\bibinfo
  {journal} {J. Fluid Mech.}\ }\textbf {\bibinfo {volume} {668}},\ \bibinfo
  {pages} {480} (\bibinfo {year} {2011})}\BibitemShut {NoStop}%
\bibitem [{\citenamefont {Zimmerman}\ \emph {et~al.}(2011)\citenamefont
  {Zimmerman}, \citenamefont {Triana},\ and\ \citenamefont
  {Lathrop}}]{Zimmerman2011}%
  \BibitemOpen
  \bibfield  {author} {\bibinfo {author} {\bibfnamefont {D.~S.}\ \bibnamefont
  {Zimmerman}}, \bibinfo {author} {\bibfnamefont {S.~A.}\ \bibnamefont
  {Triana}}, \ and\ \bibinfo {author} {\bibfnamefont {D.~P.}\ \bibnamefont
  {Lathrop}},\ }\href {\doibase 10.1063/1.3593465} {\bibfield  {journal}
  {\bibinfo  {journal} {Phys. Fluids}\ }\textbf {\bibinfo {volume} {23}},\
  \bibinfo {pages} {065104} (\bibinfo {year} {2011})}\BibitemShut {NoStop}%
\bibitem [{\citenamefont {Douady}\ \emph {et~al.}(1991)\citenamefont {Douady},
  \citenamefont {Couder},\ and\ \citenamefont {Brachet}}]{Douady1991}%
  \BibitemOpen
  \bibfield  {author} {\bibinfo {author} {\bibfnamefont {S.}~\bibnamefont
  {Douady}}, \bibinfo {author} {\bibfnamefont {Y.}~\bibnamefont {Couder}}, \
  and\ \bibinfo {author} {\bibfnamefont {M.~E.}\ \bibnamefont {Brachet}},\
  }\href {\doibase 10.1103/PhysRevLett.67.983} {\bibfield  {journal} {\bibinfo
  {journal} {Phys. Rev. Lett.}\ }\textbf {\bibinfo {volume} {67}},\ \bibinfo
  {pages} {983} (\bibinfo {year} {1991})}\BibitemShut {NoStop}%
\bibitem [{\citenamefont {Ravelet}\ \emph {et~al.}(2004)\citenamefont
  {Ravelet}, \citenamefont {Mari{\'{e}}}, \citenamefont {Chiffaudel},\ and\
  \citenamefont {Daviaud}}]{Ravelet2004}%
  \BibitemOpen
  \bibfield  {author} {\bibinfo {author} {\bibfnamefont {F.}~\bibnamefont
  {Ravelet}}, \bibinfo {author} {\bibfnamefont {L.}~\bibnamefont
  {Mari{\'{e}}}}, \bibinfo {author} {\bibfnamefont {A.}~\bibnamefont
  {Chiffaudel}}, \ and\ \bibinfo {author} {\bibfnamefont {F.}~\bibnamefont
  {Daviaud}},\ }\href {\doibase 10.1103/PhysRevLett.93.164501} {\bibfield
  {journal} {\bibinfo  {journal} {Phys. Rev. Lett.}\ }\textbf {\bibinfo
  {volume} {93}},\ \bibinfo {pages} {164501} (\bibinfo {year}
  {2004})}\BibitemShut {NoStop}%
\bibitem [{\citenamefont {De~La~Torre}\ and\ \citenamefont
  {Burguete}(2007)}]{bib:deLaTorre2007_PRL}%
  \BibitemOpen
  \bibfield  {author} {\bibinfo {author} {\bibfnamefont {A.}~\bibnamefont
  {De~La~Torre}}\ and\ \bibinfo {author} {\bibfnamefont {J.}~\bibnamefont
  {Burguete}},\ }\href {\doibase 10.1103/PhysRevLett.99.054101} {\bibfield
  {journal} {\bibinfo  {journal} {Phys. Rev. Lett.}\ }\textbf {\bibinfo
  {volume} {99}},\ \bibinfo {pages} {054101} (\bibinfo {year}
  {2007})}\BibitemShut {NoStop}%
\bibitem [{\citenamefont {Tian}\ \emph {et~al.}(2001)\citenamefont {Tian},
  \citenamefont {Weeks}, \citenamefont {Ide}, \citenamefont {Urbach},
  \citenamefont {Baroud}, \citenamefont {Ghil},\ and\ \citenamefont
  {Swinney}}]{tian2001}%
  \BibitemOpen
  \bibfield  {author} {\bibinfo {author} {\bibfnamefont {Y.}~\bibnamefont
  {Tian}}, \bibinfo {author} {\bibfnamefont {E.~R.}\ \bibnamefont {Weeks}},
  \bibinfo {author} {\bibfnamefont {K.}~\bibnamefont {Ide}}, \bibinfo {author}
  {\bibfnamefont {J.~S.}\ \bibnamefont {Urbach}}, \bibinfo {author}
  {\bibfnamefont {C.~N.}\ \bibnamefont {Baroud}}, \bibinfo {author}
  {\bibfnamefont {M.}~\bibnamefont {Ghil}}, \ and\ \bibinfo {author}
  {\bibfnamefont {H.~L.}\ \bibnamefont {Swinney}},\ }\href {\doibase
  10.1017/S0022112001004372} {\bibfield  {journal} {\bibinfo  {journal} {J.
  Fluid Mech.}\ }\textbf {\bibinfo {volume} {438}},\ \bibinfo {pages}
  {129–157} (\bibinfo {year} {2001})}\BibitemShut {NoStop}%
\bibitem [{\citenamefont {Faranda}\ \emph {et~al.}(2014)\citenamefont
  {Faranda}, \citenamefont {Bourgoin}, \citenamefont {Miralles}, \citenamefont
  {Odier}, \citenamefont {Pinton}, \citenamefont {Plihon}, \citenamefont
  {Daviaud},\ and\ \citenamefont {Dubrulle}}]{Faranda2014}%
  \BibitemOpen
  \bibfield  {author} {\bibinfo {author} {\bibfnamefont {D.}~\bibnamefont
  {Faranda}}, \bibinfo {author} {\bibfnamefont {M.}~\bibnamefont {Bourgoin}},
  \bibinfo {author} {\bibfnamefont {S.}~\bibnamefont {Miralles}}, \bibinfo
  {author} {\bibfnamefont {P.}~\bibnamefont {Odier}}, \bibinfo {author}
  {\bibfnamefont {J.~F.}\ \bibnamefont {Pinton}}, \bibinfo {author}
  {\bibfnamefont {N.}~\bibnamefont {Plihon}}, \bibinfo {author} {\bibfnamefont
  {F.}~\bibnamefont {Daviaud}}, \ and\ \bibinfo {author} {\bibfnamefont
  {B.}~\bibnamefont {Dubrulle}},\ }\href {\doibase
  10.1088/1367-2630/16/8/083001} {\bibfield  {journal} {\bibinfo  {journal}
  {New J. Phys.}\ }\textbf {\bibinfo {volume} {16}},\ \bibinfo {pages} {083001}
  (\bibinfo {year} {2014})}\BibitemShut {NoStop}%
\bibitem [{\citenamefont {Schmeits}\ and\ \citenamefont
  {Dijkstra}(2001)}]{Schmeits2001}%
  \BibitemOpen
  \bibfield  {author} {\bibinfo {author} {\bibfnamefont {M.~J.}\ \bibnamefont
  {Schmeits}}\ and\ \bibinfo {author} {\bibfnamefont {H.~A.}\ \bibnamefont
  {Dijkstra}},\ }\href {\doibase
  10.1175/1520-0485(2001)031<3435:BBOTKA>2.0.CO;2} {\bibfield  {journal}
  {\bibinfo  {journal} {J. Phys. Oceanogr.}\ }\textbf {\bibinfo {volume}
  {31}},\ \bibinfo {pages} {3435} (\bibinfo {year} {2001})}\BibitemShut
  {NoStop}%
\bibitem [{\citenamefont {Baldwin}\ \emph {et~al.}(2001)\citenamefont
  {Baldwin}, \citenamefont {Gray}, \citenamefont {Dunkerton}, \citenamefont
  {Hamilton}, \citenamefont {Haynes}, \citenamefont {Randel}, \citenamefont
  {Holton}, \citenamefont {Alexander}, \citenamefont {Hirota}, \citenamefont
  {Horinouchi}, \citenamefont {Jones}, \citenamefont {Kinnersley},
  \citenamefont {Marquardt}, \citenamefont {Sato},\ and\ \citenamefont
  {Takahashi}}]{Baldwin2001}%
  \BibitemOpen
  \bibfield  {author} {\bibinfo {author} {\bibfnamefont {M.~P.}\ \bibnamefont
  {Baldwin}}, \bibinfo {author} {\bibfnamefont {L.~J.}\ \bibnamefont {Gray}},
  \bibinfo {author} {\bibfnamefont {T.~J.}\ \bibnamefont {Dunkerton}}, \bibinfo
  {author} {\bibfnamefont {K.}~\bibnamefont {Hamilton}}, \bibinfo {author}
  {\bibfnamefont {P.~H.}\ \bibnamefont {Haynes}}, \bibinfo {author}
  {\bibfnamefont {W.~J.}\ \bibnamefont {Randel}}, \bibinfo {author}
  {\bibfnamefont {J.~R.}\ \bibnamefont {Holton}}, \bibinfo {author}
  {\bibfnamefont {M.~J.}\ \bibnamefont {Alexander}}, \bibinfo {author}
  {\bibfnamefont {I.}~\bibnamefont {Hirota}}, \bibinfo {author} {\bibfnamefont
  {T.}~\bibnamefont {Horinouchi}}, \bibinfo {author} {\bibfnamefont {D.~B.~A.}\
  \bibnamefont {Jones}}, \bibinfo {author} {\bibfnamefont {J.~S.}\ \bibnamefont
  {Kinnersley}}, \bibinfo {author} {\bibfnamefont {C.}~\bibnamefont
  {Marquardt}}, \bibinfo {author} {\bibfnamefont {K.}~\bibnamefont {Sato}}, \
  and\ \bibinfo {author} {\bibfnamefont {M.}~\bibnamefont {Takahashi}},\ }\href
  {\doibase 10.1029/1999RG000073} {\bibfield  {journal} {\bibinfo  {journal}
  {Rev. Geophys.}\ }\textbf {\bibinfo {volume} {39}},\ \bibinfo {pages} {179}
  (\bibinfo {year} {2001})}\BibitemShut {NoStop}%
\bibitem [{\citenamefont {Weeks}\ \emph {et~al.}(1997)\citenamefont {Weeks},
  \citenamefont {Tian}, \citenamefont {Urbach}, \citenamefont {Ide},
  \citenamefont {Swinney},\ and\ \citenamefont {Ghil}}]{bib:weeks1997_Science}%
  \BibitemOpen
  \bibfield  {author} {\bibinfo {author} {\bibfnamefont {E.~R.}\ \bibnamefont
  {Weeks}}, \bibinfo {author} {\bibfnamefont {Y.}~\bibnamefont {Tian}},
  \bibinfo {author} {\bibfnamefont {J.~S.}\ \bibnamefont {Urbach}}, \bibinfo
  {author} {\bibfnamefont {K.}~\bibnamefont {Ide}}, \bibinfo {author}
  {\bibfnamefont {H.~L.}\ \bibnamefont {Swinney}}, \ and\ \bibinfo {author}
  {\bibfnamefont {M.}~\bibnamefont {Ghil}},\ }\href {\doibase
  10.1126/science.278.5343.1598} {\bibfield  {journal} {\bibinfo  {journal}
  {Science}\ }\textbf {\bibinfo {volume} {278}},\ \bibinfo {pages} {1598}
  (\bibinfo {year} {1997})}\BibitemShut {NoStop}%
\bibitem [{\citenamefont {Ahmed}\ \emph {et~al.}(1984)\citenamefont {Ahmed},
  \citenamefont {Ramm},\ and\ \citenamefont {Faltin}}]{Ahmed1984}%
  \BibitemOpen
  \bibfield  {author} {\bibinfo {author} {\bibfnamefont {S.}~\bibnamefont
  {Ahmed}}, \bibinfo {author} {\bibfnamefont {G.}~\bibnamefont {Ramm}}, \ and\
  \bibinfo {author} {\bibfnamefont {G.}~\bibnamefont {Faltin}},\ }in\ \href
  {\doibase 10.4271/840300} {\emph {\bibinfo {booktitle} {SAE Technical
  Paper}}}\ (\bibinfo  {publisher} {SAE International},\ \bibinfo {year}
  {1984})\BibitemShut {NoStop}%
\bibitem [{\citenamefont {Goman}\ \emph {et~al.}(1985)\citenamefont {Goman},
  \citenamefont {Zakharov},\ and\ \citenamefont {Khrabrov}}]{Goman1985}%
  \BibitemOpen
  \bibfield  {author} {\bibinfo {author} {\bibfnamefont {M.~G.}\ \bibnamefont
  {Goman}}, \bibinfo {author} {\bibfnamefont {S.~B.}\ \bibnamefont {Zakharov}},
  \ and\ \bibinfo {author} {\bibfnamefont {A.~N.}\ \bibnamefont {Khrabrov}},\
  }\href@noop {} {\emph {\bibinfo {title} {Dokl. Akad. Nauk SSSR}}},\ \bibinfo
  {type} {Tech. Rep.}\ \bibinfo {number} {1}\ (\bibinfo {year}
  {1985})\BibitemShut {NoStop}%
\bibitem [{\citenamefont {Cadot}\ \emph {et~al.}(2015)\citenamefont {Cadot},
  \citenamefont {Evrard},\ and\ \citenamefont {Pastur}}]{Cadot2015}%
  \BibitemOpen
  \bibfield  {author} {\bibinfo {author} {\bibfnamefont {O.}~\bibnamefont
  {Cadot}}, \bibinfo {author} {\bibfnamefont {A.}~\bibnamefont {Evrard}}, \
  and\ \bibinfo {author} {\bibfnamefont {L.}~\bibnamefont {Pastur}},\ }\href
  {\doibase 10.1103/PhysRevE.91.063005} {\bibfield  {journal} {\bibinfo
  {journal} {Phys. Rev. E}\ }\textbf {\bibinfo {volume} {91}},\ \bibinfo
  {pages} {063005} (\bibinfo {year} {2015})}\BibitemShut {NoStop}%
\bibitem [{\citenamefont {Evrard}\ \emph {et~al.}(2016)\citenamefont {Evrard},
  \citenamefont {Cadot}, \citenamefont {Herbert}, \citenamefont {Ricot},
  \citenamefont {Vigneron},\ and\ \citenamefont {D{\'{e}}lery}}]{Evrard2016}%
  \BibitemOpen
  \bibfield  {author} {\bibinfo {author} {\bibfnamefont {A.}~\bibnamefont
  {Evrard}}, \bibinfo {author} {\bibfnamefont {O.}~\bibnamefont {Cadot}},
  \bibinfo {author} {\bibfnamefont {V.}~\bibnamefont {Herbert}}, \bibinfo
  {author} {\bibfnamefont {D.}~\bibnamefont {Ricot}}, \bibinfo {author}
  {\bibfnamefont {R.}~\bibnamefont {Vigneron}}, \ and\ \bibinfo {author}
  {\bibfnamefont {J.}~\bibnamefont {D{\'{e}}lery}},\ }\href {\doibase
  10.1016/j.jfluidstructs.2015.12.001} {\bibfield  {journal} {\bibinfo
  {journal} {J. Fluid Struct.}\ }\textbf {\bibinfo {volume} {61}},\ \bibinfo
  {pages} {99} (\bibinfo {year} {2016})}\BibitemShut {NoStop}%
\bibitem [{\citenamefont {Bonnavion}\ and\ \citenamefont
  {Cadot}(2018)}]{Bonnavion2018}%
  \BibitemOpen
  \bibfield  {author} {\bibinfo {author} {\bibfnamefont {G.}~\bibnamefont
  {Bonnavion}}\ and\ \bibinfo {author} {\bibfnamefont {O.}~\bibnamefont
  {Cadot}},\ }\href {\doibase 10.1017/jfm.2018.630} {\bibfield  {journal}
  {\bibinfo  {journal} {J. Fluid Mech.}\ }\textbf {\bibinfo {volume} {854}},\
  \bibinfo {pages} {196} (\bibinfo {year} {2018})}\BibitemShut {NoStop}%
\bibitem [{\citenamefont {Haffner}\ \emph {et~al.}(2020)\citenamefont
  {Haffner}, \citenamefont {Bor{\'{e}}e}, \citenamefont {Spohn},\ and\
  \citenamefont {Castelain}}]{bib:haffner2020_JFM}%
  \BibitemOpen
  \bibfield  {author} {\bibinfo {author} {\bibfnamefont {Y.}~\bibnamefont
  {Haffner}}, \bibinfo {author} {\bibfnamefont {J.}~\bibnamefont
  {Bor{\'{e}}e}}, \bibinfo {author} {\bibfnamefont {A.}~\bibnamefont {Spohn}},
  \ and\ \bibinfo {author} {\bibfnamefont {T.}~\bibnamefont {Castelain}},\
  }\href {\doibase 10.1017/jfm.2020.275} {\bibfield  {journal} {\bibinfo
  {journal} {Journal of Fluid Mechanics}\ }\textbf {\bibinfo {volume} {894}},\
  \bibinfo {pages} {A14} (\bibinfo {year} {2020})}\BibitemShut {NoStop}%
\bibitem [{\citenamefont {Hof}\ \emph {et~al.}(2006)\citenamefont {Hof},
  \citenamefont {Westerweel}, \citenamefont {Schneider},\ and\ \citenamefont
  {Eckhardt}}]{Hof2006}%
  \BibitemOpen
  \bibfield  {author} {\bibinfo {author} {\bibfnamefont {B.}~\bibnamefont
  {Hof}}, \bibinfo {author} {\bibfnamefont {J.}~\bibnamefont {Westerweel}},
  \bibinfo {author} {\bibfnamefont {T.~M.}\ \bibnamefont {Schneider}}, \ and\
  \bibinfo {author} {\bibfnamefont {B.}~\bibnamefont {Eckhardt}},\ }\href
  {\doibase 10.1038/nature05089} {\bibfield  {journal} {\bibinfo  {journal}
  {Nature}\ }\textbf {\bibinfo {volume} {443}},\ \bibinfo {pages} {59}
  (\bibinfo {year} {2006})}\BibitemShut {NoStop}%
\bibitem [{\citenamefont {Robert}\ and\ \citenamefont
  {Sommeria}(1991)}]{bib:robert1991_JFM}%
  \BibitemOpen
  \bibfield  {author} {\bibinfo {author} {\bibfnamefont {R.}~\bibnamefont
  {Robert}}\ and\ \bibinfo {author} {\bibfnamefont {J.}~\bibnamefont
  {Sommeria}},\ }\href {\doibase 10.1017/S0022112091003038} {\bibfield
  {journal} {\bibinfo  {journal} {J. Fluid Mech.}\ }\textbf {\bibinfo {volume}
  {229}},\ \bibinfo {pages} {291} (\bibinfo {year} {1991})}\BibitemShut
  {NoStop}%
\bibitem [{\citenamefont {Miller}\ \emph {et~al.}(1992)\citenamefont {Miller},
  \citenamefont {Weichman},\ and\ \citenamefont {Cross}}]{bib:miller1992_PRA}%
  \BibitemOpen
  \bibfield  {author} {\bibinfo {author} {\bibfnamefont {J.}~\bibnamefont
  {Miller}}, \bibinfo {author} {\bibfnamefont {B.}~\bibnamefont {Weichman}}, \
  and\ \bibinfo {author} {\bibfnamefont {M.~C.}\ \bibnamefont {Cross}},\ }\href
  {\doibase 10.1103/PhysRevA.45.2328} {\bibfield  {journal} {\bibinfo
  {journal} {Phys. Rev. A}\ }\textbf {\bibinfo {volume} {45}},\ \bibinfo
  {pages} {2328} (\bibinfo {year} {1992})}\BibitemShut {NoStop}%
\bibitem [{\citenamefont {Naso}\ \emph {et~al.}(2010)\citenamefont {Naso},
  \citenamefont {Monchaux},\ and\ \citenamefont {Chavanis}}]{bib:naso2010_PRE}%
  \BibitemOpen
  \bibfield  {author} {\bibinfo {author} {\bibfnamefont {A.}~\bibnamefont
  {Naso}}, \bibinfo {author} {\bibfnamefont {R.}~\bibnamefont {Monchaux}}, \
  and\ \bibinfo {author} {\bibfnamefont {P.-h.}\ \bibnamefont {Chavanis}},\
  }\href {\doibase 10.1103/PhysRevE.81.066318} {\bibfield  {journal} {\bibinfo
  {journal} {Phys. Rev. E}\ }\textbf {\bibinfo {volume} {81}},\ \bibinfo
  {pages} {066318} (\bibinfo {year} {2010})}\BibitemShut {NoStop}%
\bibitem [{\citenamefont {Bouchet}\ and\ \citenamefont
  {Venaille}(2012)}]{Bouchet2012}%
  \BibitemOpen
  \bibfield  {author} {\bibinfo {author} {\bibfnamefont {F.}~\bibnamefont
  {Bouchet}}\ and\ \bibinfo {author} {\bibfnamefont {A.}~\bibnamefont
  {Venaille}},\ }\href {\doibase 10.1016/j.physrep.2012.02.001} {\bibfield
  {journal} {\bibinfo  {journal} {Phys. Rep.}\ }\textbf {\bibinfo {volume}
  {515}},\ \bibinfo {pages} {227} (\bibinfo {year} {2012})}\BibitemShut
  {NoStop}%
\bibitem [{\citenamefont {Vulpiani}\ \emph {et~al.}(2014)\citenamefont
  {Vulpiani}, \citenamefont {Cecconi}, \citenamefont {Cencini}, \citenamefont
  {Puglisi},\ and\ \citenamefont {Vergni}}]{bib:vulpiani2014_book}%
  \BibitemOpen
  \bibfield  {author} {\bibinfo {author} {\bibfnamefont {A.}~\bibnamefont
  {Vulpiani}}, \bibinfo {author} {\bibfnamefont {F.}~\bibnamefont {Cecconi}},
  \bibinfo {author} {\bibfnamefont {M.}~\bibnamefont {Cencini}}, \bibinfo
  {author} {\bibfnamefont {A.}~\bibnamefont {Puglisi}}, \ and\ \bibinfo
  {author} {\bibfnamefont {D.}~\bibnamefont {Vergni}},\ }\href@noop {} {\emph
  {\bibinfo {title} {{Large Deviations in Physics, The Legacy of the Law of
  Large Numbers}}}}\ (\bibinfo  {publisher} {Springer},\ \bibinfo {address}
  {Berlin},\ \bibinfo {year} {2014})\BibitemShut {NoStop}%
\bibitem [{\citenamefont {Ragone}\ \emph {et~al.}(2018)\citenamefont {Ragone},
  \citenamefont {Wouters},\ and\ \citenamefont {Bouchet}}]{Ragone2018}%
  \BibitemOpen
  \bibfield  {author} {\bibinfo {author} {\bibfnamefont {F.}~\bibnamefont
  {Ragone}}, \bibinfo {author} {\bibfnamefont {J.}~\bibnamefont {Wouters}}, \
  and\ \bibinfo {author} {\bibfnamefont {F.}~\bibnamefont {Bouchet}},\ }\href
  {\doibase 10.1073/pnas.1712645115} {\bibfield  {journal} {\bibinfo  {journal}
  {P. Natl. Acad. Sci. USA}\ }\textbf {\bibinfo {volume} {115}},\ \bibinfo
  {pages} {24} (\bibinfo {year} {2018})}\BibitemShut {NoStop}%
\bibitem [{\citenamefont {Lestang}\ \emph {et~al.}(2018)\citenamefont
  {Lestang}, \citenamefont {Ragone}, \citenamefont {Br{\'{e}}hier},
  \citenamefont {Herbert},\ and\ \citenamefont {Bouchet}}]{Lestang2018}%
  \BibitemOpen
  \bibfield  {author} {\bibinfo {author} {\bibfnamefont {T.}~\bibnamefont
  {Lestang}}, \bibinfo {author} {\bibfnamefont {F.}~\bibnamefont {Ragone}},
  \bibinfo {author} {\bibfnamefont {C.~E.}\ \bibnamefont {Br{\'{e}}hier}},
  \bibinfo {author} {\bibfnamefont {C.}~\bibnamefont {Herbert}}, \ and\
  \bibinfo {author} {\bibfnamefont {F.}~\bibnamefont {Bouchet}},\ }\href
  {\doibase 10.1088/1742-5468/aab856} {\bibfield  {journal} {\bibinfo
  {journal} {J. Stat. Mech - Theory E.}\ }\textbf {\bibinfo {volume} {2018}},\
  \bibinfo {pages} {043213} (\bibinfo {year} {2018})}\BibitemShut {NoStop}%
\bibitem [{\citenamefont {Falkovich}\ \emph {et~al.}(1996)\citenamefont
  {Falkovich}, \citenamefont {Kolokolov}, \citenamefont {Lebedev},\ and\
  \citenamefont {Migdal}}]{Falkovich1996}%
  \BibitemOpen
  \bibfield  {author} {\bibinfo {author} {\bibfnamefont {G.}~\bibnamefont
  {Falkovich}}, \bibinfo {author} {\bibfnamefont {I.}~\bibnamefont
  {Kolokolov}}, \bibinfo {author} {\bibfnamefont {V.}~\bibnamefont {Lebedev}},
  \ and\ \bibinfo {author} {\bibfnamefont {A.}~\bibnamefont {Migdal}},\ }\href
  {\doibase 10.1103/PhysRevE.54.4896} {\bibfield  {journal} {\bibinfo
  {journal} {Phys. Rev. E}\ }\textbf {\bibinfo {volume} {54}},\ \bibinfo
  {pages} {4896} (\bibinfo {year} {1996})}\BibitemShut {NoStop}%
\bibitem [{\citenamefont {Dematteis}\ \emph {et~al.}(2019)\citenamefont
  {Dematteis}, \citenamefont {Grafke}, \citenamefont {Onorato},\ and\
  \citenamefont {Vanden-Eijnden}}]{Dematteis2019}%
  \BibitemOpen
  \bibfield  {author} {\bibinfo {author} {\bibfnamefont {G.}~\bibnamefont
  {Dematteis}}, \bibinfo {author} {\bibfnamefont {T.}~\bibnamefont {Grafke}},
  \bibinfo {author} {\bibfnamefont {M.}~\bibnamefont {Onorato}}, \ and\
  \bibinfo {author} {\bibfnamefont {E.}~\bibnamefont {Vanden-Eijnden}},\ }\href
  {\doibase 10.1103/PhysRevX.9.041057} {\bibfield  {journal} {\bibinfo
  {journal} {Phys. Rev. X}\ }\textbf {\bibinfo {volume} {91}},\ \bibinfo
  {pages} {41057} (\bibinfo {year} {2019})}\BibitemShut {NoStop}%
\bibitem [{\citenamefont {Grafke}\ \emph {et~al.}(2015)\citenamefont {Grafke},
  \citenamefont {Grauer},\ and\ \citenamefont {Sch{\"{a}}fer}}]{Grafke2015}%
  \BibitemOpen
  \bibfield  {author} {\bibinfo {author} {\bibfnamefont {T.}~\bibnamefont
  {Grafke}}, \bibinfo {author} {\bibfnamefont {R.}~\bibnamefont {Grauer}}, \
  and\ \bibinfo {author} {\bibfnamefont {T.}~\bibnamefont {Sch{\"{a}}fer}},\
  }\href {\doibase 10.1088/1751-8113/48/33/333001} {\bibfield  {journal}
  {\bibinfo  {journal} {J. Phys. A - Math. Theor.}\ }\textbf {\bibinfo {volume}
  {48}},\ \bibinfo {pages} {333001} (\bibinfo {year} {2015})}\BibitemShut
  {NoStop}%
\bibitem [{\citenamefont {Reynolds}(1883)}]{Reynolds1883}%
  \BibitemOpen
  \bibfield  {author} {\bibinfo {author} {\bibfnamefont {O.}~\bibnamefont
  {Reynolds}},\ }\href {\doibase 10.1098/rspl.1883.0018} {\bibfield  {journal}
  {\bibinfo  {journal} {Proc. R. Soc. Lond.}\ }\textbf {\bibinfo {volume}
  {35}},\ \bibinfo {pages} {84} (\bibinfo {year} {1883})}\BibitemShut {NoStop}%
\bibitem [{\citenamefont {Barkley}(2016)}]{Barkley2016}%
  \BibitemOpen
  \bibfield  {author} {\bibinfo {author} {\bibfnamefont {D.}~\bibnamefont
  {Barkley}},\ }\href {\doibase 10.1017/jfm.2016.465} {\bibfield  {journal}
  {\bibinfo  {journal} {J. Fluid Mech.}\ }\textbf {\bibinfo {volume} {803}},\
  \bibinfo {pages} {P1} (\bibinfo {year} {2016})}\BibitemShut {NoStop}%
\bibitem [{\citenamefont {Avila}\ \emph {et~al.}(2011)\citenamefont {Avila},
  \citenamefont {Moxey}, \citenamefont {de~Lozar}, \citenamefont {Avila},
  \citenamefont {Barkley},\ and\ \citenamefont {Hof}}]{Avila2011}%
  \BibitemOpen
  \bibfield  {author} {\bibinfo {author} {\bibfnamefont {K.}~\bibnamefont
  {Avila}}, \bibinfo {author} {\bibfnamefont {D.}~\bibnamefont {Moxey}},
  \bibinfo {author} {\bibfnamefont {A.}~\bibnamefont {de~Lozar}}, \bibinfo
  {author} {\bibfnamefont {M.}~\bibnamefont {Avila}}, \bibinfo {author}
  {\bibfnamefont {D.}~\bibnamefont {Barkley}}, \ and\ \bibinfo {author}
  {\bibfnamefont {B.}~\bibnamefont {Hof}},\ }\href {\doibase
  10.1126/science.1203223} {\bibfield  {journal} {\bibinfo  {journal}
  {Science}\ }\textbf {\bibinfo {volume} {333}},\ \bibinfo {pages} {192}
  (\bibinfo {year} {2011})}\BibitemShut {NoStop}%
\bibitem [{\citenamefont {Goldenfeld}\ \emph {et~al.}(2010)\citenamefont
  {Goldenfeld}, \citenamefont {Guttenberg},\ and\ \citenamefont
  {Gioia}}]{Goldenfeld2010}%
  \BibitemOpen
  \bibfield  {author} {\bibinfo {author} {\bibfnamefont {N.}~\bibnamefont
  {Goldenfeld}}, \bibinfo {author} {\bibfnamefont {N.}~\bibnamefont
  {Guttenberg}}, \ and\ \bibinfo {author} {\bibfnamefont {G.}~\bibnamefont
  {Gioia}},\ }\href {\doibase 10.1103/PhysRevE.81.035304} {\bibfield  {journal}
  {\bibinfo  {journal} {Phys. Rev. E}\ }\textbf {\bibinfo {volume} {81}},\
  \bibinfo {pages} {035304} (\bibinfo {year} {2010})}\BibitemShut {NoStop}%
\bibitem [{\citenamefont {Nemoto}\ and\ \citenamefont
  {Alexakis}(2018)}]{Nemoto2018}%
  \BibitemOpen
  \bibfield  {author} {\bibinfo {author} {\bibfnamefont {T.}~\bibnamefont
  {Nemoto}}\ and\ \bibinfo {author} {\bibfnamefont {A.}~\bibnamefont
  {Alexakis}},\ }\href {\doibase 10.1103/PhysRevE.97.022207} {\bibfield
  {journal} {\bibinfo  {journal} {Phys. Rev. E}\ }\textbf {\bibinfo {volume}
  {97}},\ \bibinfo {pages} {022207} (\bibinfo {year} {2018})}\BibitemShut
  {NoStop}%
\bibitem [{\citenamefont {Schikarski}\ \emph {et~al.}(2019)\citenamefont
  {Schikarski}, \citenamefont {Trzenschiok}, \citenamefont {Peukert},\ and\
  \citenamefont {Avila}}]{Schikarski2019}%
  \BibitemOpen
  \bibfield  {author} {\bibinfo {author} {\bibfnamefont {T.}~\bibnamefont
  {Schikarski}}, \bibinfo {author} {\bibfnamefont {H.}~\bibnamefont
  {Trzenschiok}}, \bibinfo {author} {\bibfnamefont {W.}~\bibnamefont
  {Peukert}}, \ and\ \bibinfo {author} {\bibfnamefont {M.}~\bibnamefont
  {Avila}},\ }\href {\doibase 10.1039/c8re00208h} {\bibfield  {journal}
  {\bibinfo  {journal} {React. Chem. Eng.}\ }\textbf {\bibinfo {volume} {4}},\
  \bibinfo {pages} {559} (\bibinfo {year} {2019})}\BibitemShut {NoStop}%
\bibitem [{\citenamefont {Nemoto}\ and\ \citenamefont
  {Alexakis}(2020)}]{Nemoto2020}%
  \BibitemOpen
  \bibfield  {author} {\bibinfo {author} {\bibfnamefont {T.}~\bibnamefont
  {Nemoto}}\ and\ \bibinfo {author} {\bibfnamefont {A.}~\bibnamefont
  {Alexakis}},\ }\href {http://arxiv.org/abs/2005.03530} {\  (\bibinfo {year}
  {2020})},\ \Eprint {http://arxiv.org/abs/2005.03530} {arXiv:2005.03530}
  \BibitemShut {NoStop}%
\bibitem [{\citenamefont {Obligado}\ \emph {et~al.}(2013)\citenamefont
  {Obligado}, \citenamefont {Puy},\ and\ \citenamefont
  {Bourgoin}}]{Obligado2013}%
  \BibitemOpen
  \bibfield  {author} {\bibinfo {author} {\bibfnamefont {M.}~\bibnamefont
  {Obligado}}, \bibinfo {author} {\bibfnamefont {M.}~\bibnamefont {Puy}}, \
  and\ \bibinfo {author} {\bibfnamefont {M.}~\bibnamefont {Bourgoin}},\ }\href
  {\doibase 10.1017/jfm.2013.312} {\bibfield  {journal} {\bibinfo  {journal}
  {J. Fluid Mech.}\ }\textbf {\bibinfo {volume} {728}},\ \bibinfo {pages} {R2}
  (\bibinfo {year} {2013})}\BibitemShut {NoStop}%
\bibitem [{\citenamefont {Linkmann}\ and\ \citenamefont
  {Morozov}(2015)}]{Linkmann2015}%
  \BibitemOpen
  \bibfield  {author} {\bibinfo {author} {\bibfnamefont {M.~F.}\ \bibnamefont
  {Linkmann}}\ and\ \bibinfo {author} {\bibfnamefont {A.}~\bibnamefont
  {Morozov}},\ }\href {\doibase 10.1103/PhysRevLett.115.134502} {\bibfield
  {journal} {\bibinfo  {journal} {Phys. Rev. Lett.}\ }\textbf {\bibinfo
  {volume} {115}},\ \bibinfo {pages} {134502} (\bibinfo {year}
  {2015})}\BibitemShut {NoStop}%
\bibitem [{\citenamefont {Shi}\ \emph {et~al.}(2013)\citenamefont {Shi},
  \citenamefont {Avila},\ and\ \citenamefont {Hof}}]{Shi2013}%
  \BibitemOpen
  \bibfield  {author} {\bibinfo {author} {\bibfnamefont {L.}~\bibnamefont
  {Shi}}, \bibinfo {author} {\bibfnamefont {M.}~\bibnamefont {Avila}}, \ and\
  \bibinfo {author} {\bibfnamefont {B.}~\bibnamefont {Hof}},\ }\href {\doibase
  10.1103/PhysRevLett.110.204502} {\bibfield  {journal} {\bibinfo  {journal}
  {Phys. Rev. Lett.}\ }\textbf {\bibinfo {volume} {110}},\ \bibinfo {pages}
  {204502} (\bibinfo {year} {2013})}\BibitemShut {NoStop}%
\bibitem [{\citenamefont {Gom{\'{e}}}\ \emph {et~al.}(2020)\citenamefont
  {Gom{\'{e}}}, \citenamefont {Tuckerman},\ and\ \citenamefont
  {Barkley}}]{Gome2020}%
  \BibitemOpen
  \bibfield  {author} {\bibinfo {author} {\bibfnamefont {S.}~\bibnamefont
  {Gom{\'{e}}}}, \bibinfo {author} {\bibfnamefont {L.~S.}\ \bibnamefont
  {Tuckerman}}, \ and\ \bibinfo {author} {\bibfnamefont {D.}~\bibnamefont
  {Barkley}},\ }\href {\doibase 10.1103/physrevfluids.5.083905} {\bibfield
  {journal} {\bibinfo  {journal} {Phys. Rev. Fluids}\ }\textbf {\bibinfo
  {volume} {5}},\ \bibinfo {pages} {083905} (\bibinfo {year}
  {2020})}\BibitemShut {NoStop}%
\bibitem [{\citenamefont {Flachsbart}(1932)}]{Flachsbart1932}%
  \BibitemOpen
  \bibfield  {author} {\bibinfo {author} {\bibfnamefont {O.}~\bibnamefont
  {Flachsbart}},\ }in\ \href@noop {} {\emph {\bibinfo {booktitle} {Ergebnisse
  der Aerodynamischen Versuchsanstalt zu G{\"{o}}ttingen - IV. Lieferung}}}\
  (\bibinfo  {publisher} {Verlag von R. Oldenburg},\ \bibinfo {address}
  {M{\"{u}}nchen und Berlin},\ \bibinfo {year} {1932})\ pp.\ \bibinfo {pages}
  {96--100}\BibitemShut {NoStop}%
\bibitem [{\citenamefont {Ericsson}\ and\ \citenamefont
  {Reding}(1980)}]{Ericsson1980}%
  \BibitemOpen
  \bibfield  {author} {\bibinfo {author} {\bibfnamefont {L.~E.}\ \bibnamefont
  {Ericsson}}\ and\ \bibinfo {author} {\bibfnamefont {J.~P.}\ \bibnamefont
  {Reding}},\ }\href {\doibase 10.2514/3.57884} {\bibfield  {journal} {\bibinfo
   {journal} {J. Aircraft}\ }\textbf {\bibinfo {volume} {17}},\ \bibinfo
  {pages} {136} (\bibinfo {year} {1980})}\BibitemShut {NoStop}%
\bibitem [{\citenamefont {McCroskey}\ \emph {et~al.}(1976)\citenamefont
  {McCroskey}, \citenamefont {Carr},\ and\ \citenamefont
  {McAlister}}]{McCroskey1975}%
  \BibitemOpen
  \bibfield  {author} {\bibinfo {author} {\bibfnamefont {W.~J.}\ \bibnamefont
  {McCroskey}}, \bibinfo {author} {\bibfnamefont {L.~W.}\ \bibnamefont {Carr}},
  \ and\ \bibinfo {author} {\bibfnamefont {K.~W.}\ \bibnamefont {McAlister}},\
  }\href {\doibase 10.2514/6.1975-125} {\bibfield  {journal} {\bibinfo
  {journal} {AIAA J.}\ }\textbf {\bibinfo {volume} {14}},\ \bibinfo {pages}
  {57} (\bibinfo {year} {1976})}\BibitemShut {NoStop}%
\bibitem [{\citenamefont {Avila}\ \emph {et~al.}(2010)\citenamefont {Avila},
  \citenamefont {Willis},\ and\ \citenamefont {Hof}}]{Avila2010}%
  \BibitemOpen
  \bibfield  {author} {\bibinfo {author} {\bibfnamefont {M.}~\bibnamefont
  {Avila}}, \bibinfo {author} {\bibfnamefont {A.~P.}\ \bibnamefont {Willis}}, \
  and\ \bibinfo {author} {\bibfnamefont {B.}~\bibnamefont {Hof}},\ }\href
  {\doibase 10.1017/S0022112009993296} {\bibfield  {journal} {\bibinfo
  {journal} {J. Fluid Mech.}\ }\textbf {\bibinfo {volume} {646}},\ \bibinfo
  {pages} {127} (\bibinfo {year} {2010})}\BibitemShut {NoStop}%
\bibitem [{\citenamefont {Hof}\ \emph {et~al.}(2008)\citenamefont {Hof},
  \citenamefont {{De Lozar}}, \citenamefont {Kuik},\ and\ \citenamefont
  {Westerweel}}]{Hof2008}%
  \BibitemOpen
  \bibfield  {author} {\bibinfo {author} {\bibfnamefont {B.}~\bibnamefont
  {Hof}}, \bibinfo {author} {\bibfnamefont {A.}~\bibnamefont {{De Lozar}}},
  \bibinfo {author} {\bibfnamefont {D.~J.}\ \bibnamefont {Kuik}}, \ and\
  \bibinfo {author} {\bibfnamefont {J.}~\bibnamefont {Westerweel}},\ }\href
  {\doibase 10.1103/PhysRevLett.101.214501} {\bibfield  {journal} {\bibinfo
  {journal} {Phys. Rev. Lett.}\ }\textbf {\bibinfo {volume} {101}},\ \bibinfo
  {pages} {214501} (\bibinfo {year} {2008})}\BibitemShut {NoStop}%
\bibitem [{Note1()}]{Note1}%
  \BibitemOpen
  \bibinfo {note} {The mean value being $\mu +\beta \gamma $ , with $\gamma
  \simeq 0.577$ the Euler-Mascheroni constant and the standard deviation being
  $\beta \pi /\protect \sqrt {6}$. For each value of $\protect \overline
  {\theta _i}$, $\beta $ and $\mu $ have been extracted from experimental
  signals.}\BibitemShut {Stop}%
\bibitem [{\citenamefont {Gao}\ \emph {et~al.}(2018)\citenamefont {Gao},
  \citenamefont {Tao}, \citenamefont {Tian},\ and\ \citenamefont
  {Yang}}]{Gao2018}%
  \BibitemOpen
  \bibfield  {author} {\bibinfo {author} {\bibfnamefont {S.}~\bibnamefont
  {Gao}}, \bibinfo {author} {\bibfnamefont {L.}~\bibnamefont {Tao}}, \bibinfo
  {author} {\bibfnamefont {X.}~\bibnamefont {Tian}}, \ and\ \bibinfo {author}
  {\bibfnamefont {J.}~\bibnamefont {Yang}},\ }\href {\doibase
  10.1017/jfm.2018.526} {\bibfield  {journal} {\bibinfo  {journal} {J. Fluid
  Mech.}\ }\textbf {\bibinfo {volume} {851}},\ \bibinfo {pages} {687} (\bibinfo
  {year} {2018})}\BibitemShut {NoStop}%
\end{thebibliography}%
\end{document}